\RequirePackage{fix-cm}
\documentclass[smallextended]{svjour3}       % onecolumn (second format)
\smartqed  % flush right qed marks, e.g. at end of proof
\usepackage{graphicx}

\usepackage{cite}

\usepackage{amsmath}
\usepackage{amssymb}
\usepackage[colorlinks=true, citecolor=blue]{hyperref}
\usepackage{color}

\begin{document}

\title{Two topologically distinct Dirac-line semimetal phases and topological phase transitions in rhombohedrally stacked honeycomb lattices %\thanks{Grants or other notes
%about the article that should go on the front page should be
%placed here. General acknowledgments should be placed at the end of the article.}
}
%\subtitle{Do you have a subtitle?\\ If so, write it here}

\titlerunning{Two topologically distinct Dirac-line semimetal phases}        % if too long for running head

\author{T.~Hyart         \and R.~Ojaj\"arvi \and
        T.~T.~Heikkil\"a %etc.
}

%\authorrunning{Short form of author list} % if too long for running head

\institute{T. Hyart  \at
              Department of Physics and Nanoscience Center, University of Jyv\"askyl\"a, P.O. Box 35 (YFL), FI-40014 University of Jyv\"askyl\"a, Finland 
              %Tel.: +123-45-678910\\
              %Fax: +123-45-678910\\
             % \email{fauthor@example.com}           %  \\
%             \emph{Present address:} of F. Author  %  if needed
	\and
           R.~Ojaj\"arvi \at
              Department of Physics and Nanoscience Center, University of Jyv\"askyl\"a, P.O. Box 35 (YFL), FI-40014 University of Jyv\"askyl\"a, Finland
           \and
           T.~T.~Heikkil\"a \at
              Department of Physics and Nanoscience Center, University of Jyv\"askyl\"a, P.O. Box 35 (YFL), FI-40014 University of Jyv\"askyl\"a, Finland
}

\date{Received: date / Accepted: date}
% The correct dates will be entered by the editor

\maketitle

\begin{abstract}
Three-dimensional topological semimetals can support band crossings along one-dimensional curves in the momentum space (nodal lines or Dirac lines) protected by structural symmetries and topology.  We consider rhombohedrally (ABC) stacked honeycomb lattices supporting Dirac lines protected by time-reversal, inversion and spin rotation symmetries. For typical  band structure parameters there exists a pair of nodal lines in the momentum space extending through the whole Brillouin zone in the stacking direction. We show that these Dirac lines are topologically distinct from the usual Dirac lines which form closed loops inside the Brillouin zone. In particular, an energy gap can be opened only by first merging the Dirac lines going through the Brillouin zone in a pairwise manner so that they turn into closed loops inside the Brillouin zone, and then by shrinking these loops into points. We show that this kind of topological phase transition can occur in rhombohedrally stacked honeycomb lattices by tuning the ratio of the tunneling amplitudes in the directions perpendicular and parallel to the layers.  We also discuss the properties of the surface states in the different phases of the model.

%\keywords{First keyword \and Second keyword \and More}
% \PACS{PACS code1 \and PACS code2 \and more}
% \subclass{MSC code1 \and MSC code2 \and more}
\end{abstract}

\section{Introduction}
\label{intro}

Topological materials are characterized by momentum-space topological defects, topological invariants and protected surface states \cite{Volovik-book, HaKa10, Zhang-review, Schny+08, SchnyderRMP07}. The fully gapped topological phases have been classified in terms of the existence of various symmetries \cite{Schny+08, SchnyderRMP07}, and the variety of the different types of momentum space topological defects is even richer in gapless systems \cite{Volovik-book, SchnyderRMP07, Nielsen83,Volovik87, Ryu02, McClure57, Horava, Mikitik06, Mikitik08, Heikkila-Volovik11, Heikkila-flat-bands, BurkovBalents, Kim15, Fu15, Chan15, nexus, nexus_Hyart, nexus_paper, NewFermionPaper, nodalchain, Ezawa16, Horsdal17, Yan17, Ezawa17,Bzdusek, Bouhon17}. One interesting class of three-dimensional topological semimetals are the Dirac-line (nodal line) semimetals supporting band crossings along one-dimensional curves in the momentum space. These band crossings can in principle be protected by chiral symmetry \cite{Volovik-book, SchnyderRMP07, Ryu02, Heikkila-Volovik11} (often an emergent or an approximate symmetry) or the structural symmetries of the systems \cite{SchnyderRMP07, BurkovBalents, Kim15, Fu15, Chan15, nexus_Hyart, Bzdusek, Bouhon17}. 
From the viewpoint of topological materials the main question is what kind of topologically distinct Dirac-line semimetal phases can exist in the presence of the various structural symmetries. For example, there exists a class of nodal lines carrying a nontrivial $\mathbb{Z}_2$ monopole charge so that they can be created and annihilated only in pairs, whereas the nodal lines carrying a trivial $\mathbb{Z}_2$ monopole charge can be created and annihilated one by one \cite{Fu15, Bzdusek, Bouhon17}. In certain topological semimetals, such as Bernally stacked graphite, there exists multiple Dirac lines which meet and merge at certain high-symmetry lines in the momentum space forming a protected triple degeneracy point of bands called nexus \cite{nexus, nexus_Hyart, nexus_paper}.

\begin{figure*}[b]
  \includegraphics[width=0.9\textwidth]{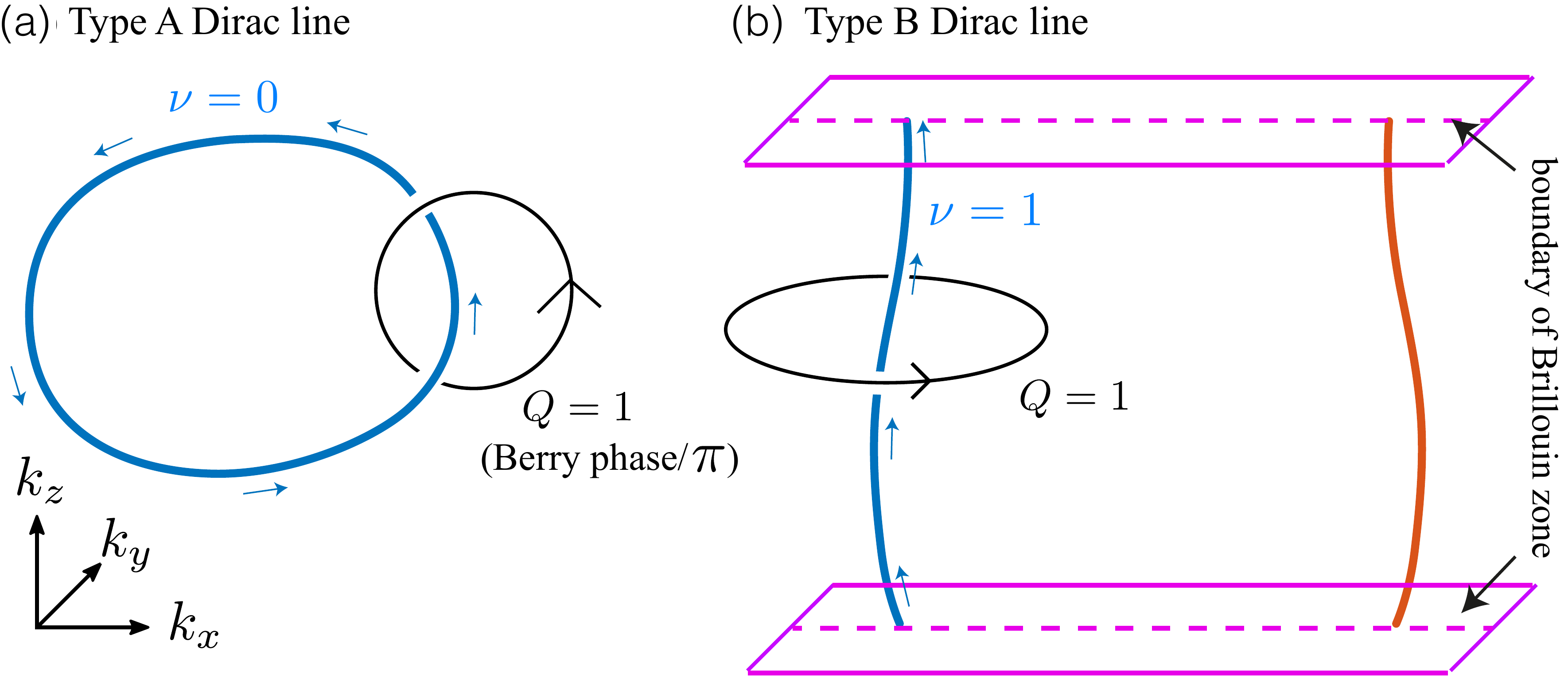}
\caption{Two different types of Dirac-line semimetal phases in the presence of SU(2) spin rotation, time-reversal and inversion symmetries. (a) In Type A Dirac-line semimetals the nodal lines are closed loops fully inside the Brillouin zone (blue line). Thus, they go through the full Brillouin zone $\nu=0$ times. They carry a topological charge $Q=1$ defined with the help of a Berry phase for a path going around the Dirac line. Type A Dirac lines can be gapped one by one.  (b) In Type B Dirac-line semimetals the nodal lines extend through the whole Brillouin zone in one of the directions (blue and red curves). Thus, they go through the full Brillouin zone $\nu=1$ times. They carry a topological charge $Q=1$ which can be defined with the help of Berry phase, but now the radius of the closed path going around the Dirac line can be taken arbitrary large (as long as it does not go around another Dirac line). Therefore, it is possible to define a topological charge for each Type B Dirac line as an integral over a closed surface [Eq.~(\ref{monopole-charge})] describing their monopole-like nature: Type B Dirac lines can be gapped only by first merging them in a pairwise manner.}
\label{fig:typeA_vs_typeB}       % Give a unique label
\end{figure*}

In this paper we study Dirac lines protected by time-reversal, inversion and spin rotation symmetries with the help of a general model for rhombohedrally (ABC) stacked honeycomb lattices. We show that these Dirac lines can form closed loops inside the Brillouin zone (Type A Dirac line in Fig.~\ref{fig:typeA_vs_typeB}) \cite{SchnyderRMP07, BurkovBalents, Kim15, Fu15, Chan15}  or they can extend through the whole Brillouin zone in one of the directions in the momentum space (Type B Dirac line in Fig.~\ref{fig:typeA_vs_typeB}) \cite{Heikkila-Volovik11} depending on the ratio of intra- and interlayer tunneling amplitudes. We show that Type A and Type B Dirac lines are topologically distinct. Namely, it is possible define a topological invariant $\nu$ by counting how many times the Dirac line goes through the whole Brillouin zone, and $\nu=0$ ($\nu=1$) for Type A (Type B) Dirac lines (see Fig.~\ref{fig:typeA_vs_typeB}). Since the Brillouin zone is a torus, $\nu$ essentially describes the winding of the Dirac line around this torus. We show that this topological difference has important consequences. Namely, it allows defining a topological charge for Type B Dirac lines describing their monopole-like nature (different from the monopole charge proposed in \cite{Fu15}): Type B Dirac lines are robust nodal lines which must always come in pairs. Therefore, Type A Dirac lines can be created and annihilated individually by shrinking them to a point but Type B Dirac lines can be gapped only by first merging them in a pairwise manner so that they become Type A Dirac lines. We discuss the nature of this kind of topological phase transition in rhombohedrally stacked honeycomb lattices and depict the momentum-space structure of the surface states for the different phases of the model. We point out that 
the monopole-like nature and robustness of the Type B Dirac lines have been discussed also in Refs.~\cite{Bzdusek, Bouhon17} from different perspectives.

\section{Model, symmetries and topological invariants}
\label{sec:model}

\begin{figure}
\begin{center}
  \includegraphics[width=0.72\textwidth]{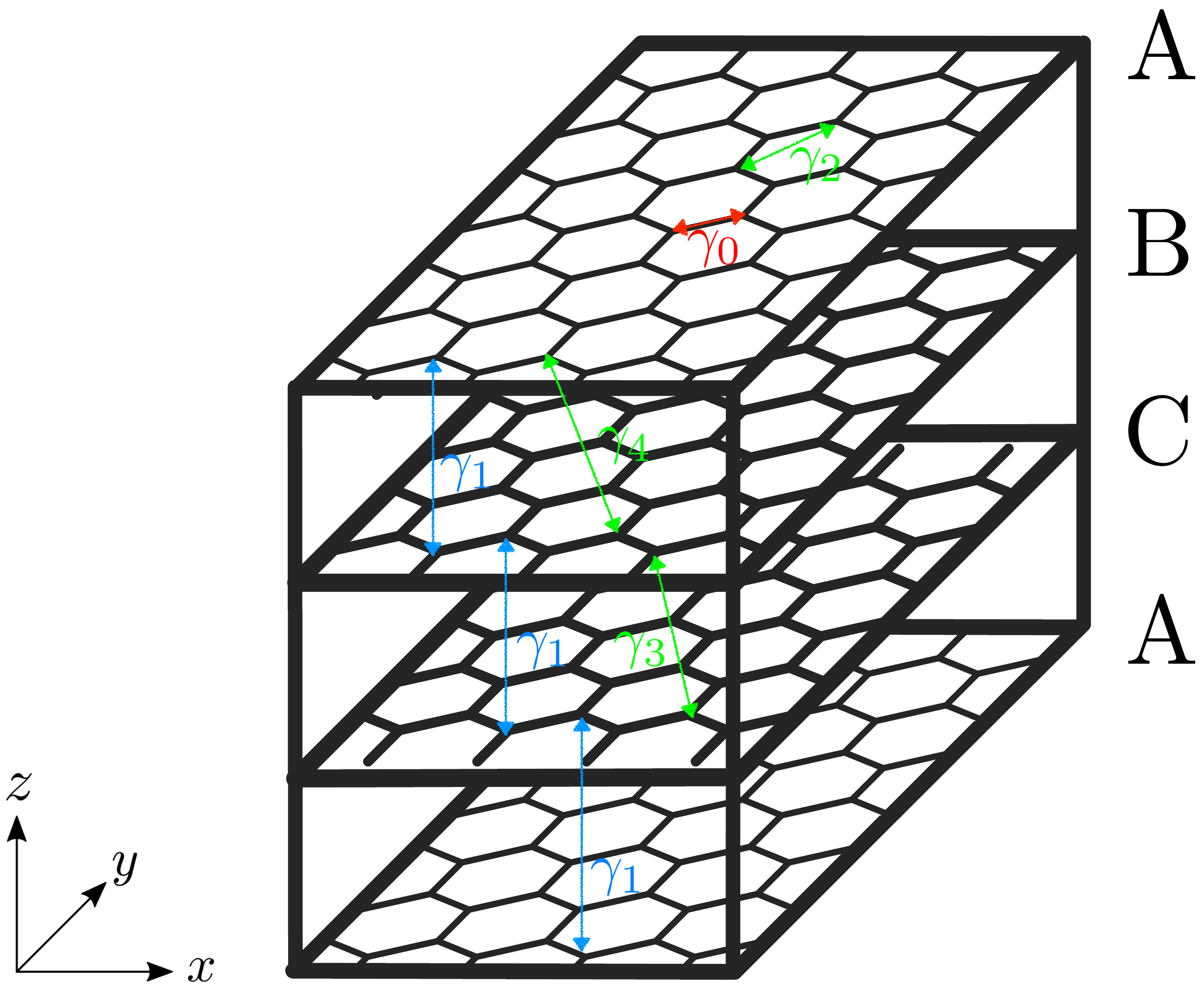}
  \end{center}
\caption{Illustration of rhombohedrally stacked honeycomb lattices. The hopping amplitude $\gamma_0$ describes the nearest-neighbor hopping inside the layers. In rhombohedral stacking the layers are stacked in such a way that the nearest-neighbor interlayer hopping amplitudes $\gamma_1$ always connect one type of sublattice atom in the lower layer to the other type of sublattice atom in the upper layer. The further neighbor hoppings $\gamma_2$, $\gamma_3$ and $\gamma_4$ are also shown. The period in the stacking direction is three times the distance between the layers and therefore rhombohedral stacking is often called ABC stacking.}
\label{fig:rhombostacking}       
\end{figure}

The rhombohedral stacking of honeycomb lattices is illustrated in Fig.~\ref{fig:rhombostacking}. The tight-binding model for such kind of three-dimensional system [in the sublattice space] can be written as 
\begin{eqnarray}
H(\mathbf{k})&=& \begin{pmatrix}
   \Theta(\mathbf{k})    &  \Phi(\mathbf{k})\\
 \Phi^*(\mathbf{k})    &   \Theta(\mathbf{k})   
\end{pmatrix},  \label{full-H}  \\
 \Phi(\mathbf{k})&=&- \gamma_0 \sum_i e^{i \vec{\delta}_i \cdot \vec{k}}-\gamma_1 e^{i k_z} - \gamma_3 e^{-i k_z} \sum_i e^{-i \vec{\delta}_i \cdot \vec{k}} \nonumber \\
 \Theta(\mathbf{k})&=&- \gamma_2 \sum_i e^{i \vec{n}_i \cdot \vec{k}}- \gamma_4 \bigg[ e^{ik_z}\sum_i e^{-i \vec{\delta}_i \cdot \vec{k}} + e^{-ik_z}\sum_i e^{i \vec{\delta}_i \cdot \vec{k}} \bigg] \nonumber
\end{eqnarray}
The hopping parameters $\gamma_i$  are illustrated in Fig.~\ref{fig:rhombostacking}.  The nearest-neighbor vectors $\vec{\delta}_i$ inside the layers (connecting different types of sublattice atoms) are normalized so that the vectors $\vec{n}_i$ connecting neighboring unit cells inside the layers (i.e. connecting the same type of sublattice atoms) have unit length. In $z$-direction we use the spacing between the layers as the unit length. 

The important symmetries of the model are the lattice translation symmetries (guaranteeing that $\vec{k}$ is a good quantum number), SU(2) spin rotation symmetry [so that we do not need to include spin degree of freedom into the Hamiltonian (\ref{full-H})], time-reversal symmetry 
\begin{equation}
H^*(-\mathbf{k})=H(\mathbf{k})
\end{equation}
and inversion symmetry
\begin{equation}
\sigma_x H(-\mathbf{k}) \sigma_x=H(\mathbf{k}).
\end{equation}
In many materials, such as graphite, spin-orbit coupling is negligible, and therefore the assumption about SU(2) spin rotation symmetry is well justified. The other symmetries are just structural symmetries of the system and therefore they are satisfied independently on which hopping processes are included in the Hamiltonian. They could be broken only via a spontaneous symmetry breaking occurring due to interactions between the fermions. 

It is useful to point out that if the further neighbor hoppings are neglected $\gamma_2=\gamma_4=0$, the system supports also a chiral symmetry 
\begin{equation}
\sigma_z H(\mathbf{k}) \sigma_z = -H(\mathbf{k}). \label{chiral-symmetry}
\end{equation}
This symmetry is valid as a good approximation in many systems, such as graphite, and it is useful in understanding the surface state spectrum of the model (see Sec.~\ref{sec:surface_states}). The chiral symmetry can in principle also protect the existence of the Dirac lines \cite{Heikkila-Volovik11}. However, in this paper we show that this approximate chiral symmetry is not necessary for the existence of the  Dirac lines. The Dirac lines in this model are in fact much more robust because they are stabilized by the structural symmetries of the system.  

Due to the existence of SU(2), time-reversal and inversion symmetries $\sigma_x H^*(\mathbf{k}) \sigma_x=H(\mathbf{k})$, and therefore for any closed path in the momentum space the Berry phase is quantized to be $\phi=0$ or $\phi=\pi$ \cite{comment}.   Moreover, it is easy to show that $\phi=\pi$ if and only if the path goes around a Dirac line. Therefore, the Dirac lines in this symmetry class carry a $\mathbb{Z}_2$ topological charge, which can be defined as $Q \equiv \phi/\pi$, where $\phi$ can be chosen to be the Berry phase for any path that goes around the Dirac line (see Fig.~\ref{fig:typeA_vs_typeB}). Thus, the Dirac lines are stable against small perturbations. If the Dirac line forms a closed loop within the Brillouin zone (Type A Dirac line),  it can be smoothly annihilated only by first shrinking it to a point so that we can no longer define a path going around it.  However, there are no restrictions concerning the number of Type A Dirac lines because they can be gapped one by one in this way.

Interestingly, there exists also topologically different type of Dirac lines in this symmetry class. Intuitively, this is easy to understand because we can visualize the Brillouin zone as a three-dimensional torus and we can define three different topological invariants for each Dirac line by counting how many times the Dirac line winds around the torus in different directions. Here, we concentrate on this kind of topological invariant $\nu$ corresponding to the stacking direction $k_z$.  For the Type A Dirac lines, which are closed loops fully inside the Brillouin zone, $\nu=0$. The other topologically different Dirac lines extend through the full Brillouin zone in $k_z$ direction, so that $\nu \ne 0$ (see Fig.~\ref{fig:typeA_vs_typeB}). It turns out that these Dirac lines have different properties than the ones with $\nu=0$. In order to understand these differences we consider the Dirac lines with $\nu=1$ shown in Fig.~\ref{fig:typeA_vs_typeB}(b). Similarly as above the Berry phase $\phi=\pi$ for any closed path going around the Dirac line. For each value of $k_z$ we can now choose a path  in the $(k_x, k_y)$-plane with arbitrary large radius (as long as it does not go around another Dirac line), and we can express the corresponding Berry phase as a line integral around that path $\phi=\oint_l d\mathbf{k} \cdot \vec{\cal{A}}(\mathbf{k})$, where $\vec{\cal{A}}(\mathbf{k})$ is the Berry connection \cite{comment}. Since the line integral is $\phi=\pi$ independently on $k_z$, we can define a new topological charge  
\begin{equation}
Q_M=\frac{1}{2 \pi^2} \int_{-\pi}^\pi dk_z \oint_l d\mathbf{k} \cdot \cal{A}(\mathbf{k}), \label{monopole-charge}
\end{equation}
so that for $\nu=1$ this topological charge is quantized to $Q_M=1$. The difference to the earlier topological charge $Q$ is that $Q_M$ is expressed as a surface integral over a closed surface which encloses the whole Dirac line, and this surface can be taken arbitrarily far away from the Dirac line as long as it does not enclose any parts of other Dirac lines. (The surface is closed if one considers the momentum space as a torus so that $k_z= \pm \pi$ are the same point within the torus.) Therefore, $Q_M$ can be considered as a topological charge calculated over a closed surface, which is different from the monopole charge proposed in Ref.~\cite{Fu15}.  Furthermore, $Q_M$ distinguishes Type A and Type B Dirac lines since for Type A Dirac lines $Q_M=0$. Thus, the topological charge $Q_M$ can be used in the identification and search of Type B Dirac lines. More generally, it is related to $\nu$ because $Q_M$ = $\nu$ mod 2. 

If we now take the path in calculation of the Berry phase to go around the full Brillouin zone in the $(k_x,k_y)$-plane, we find that the surface integral is always necessarily zero. Therefore, it is impossible to have only a single Dirac line with $\nu=1$ in the Brillouin zone. This means that the existence of the charge $Q_M$ enforces the Dirac lines to come in pairs.  Therefore, these Dirac lines cannot be created and annihilated individually, and the only way to open a gap in a Type B Dirac-line semimetal is to first merge the Dirac lines in a pairwise manner. (More generally, in this symmetry class the Dirac lines with odd $\nu$ must always come in pairs \cite{comment2}.) In Sec.~\ref{phase_diagram} we illustrate this type of topological phase transition taking place as a function of $\gamma_1/\gamma_0$ in the model for rhombohedrally stacked honeycomb lattices [Eq.~(\ref{full-H})].

\section{Phase diagram}
\label{phase_diagram}

\begin{figure}
\begin{center}
   \includegraphics[width=0.99\textwidth]{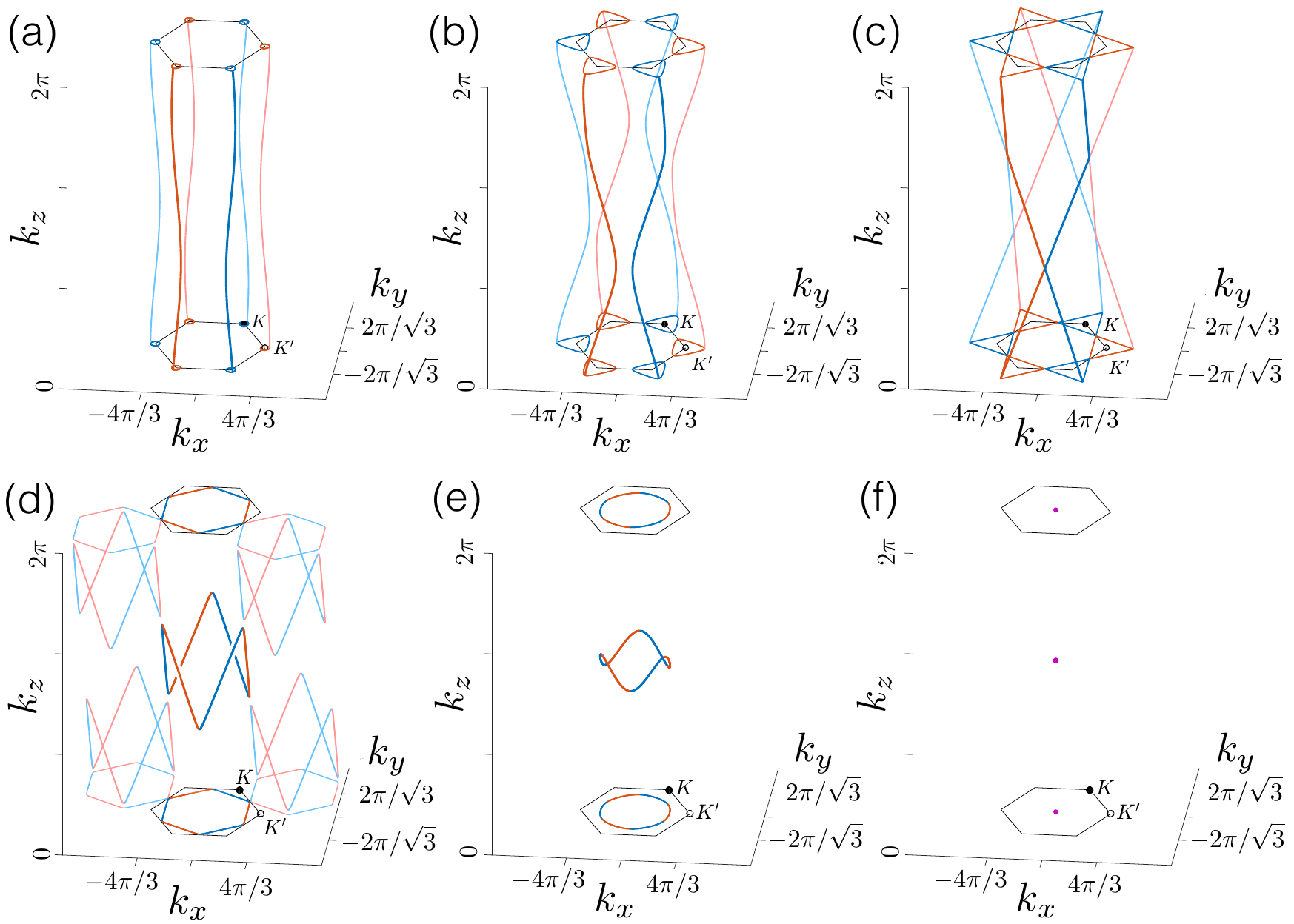}
  \end{center}
\caption{Evolution of the Dirac lines in the model for rhombohedrally stacked honeycomb lattices [Eq.~(\ref{full-H})] with increasing $\gamma_1/\gamma_0$: (a) $\gamma_1/\gamma_0=0.3$, (b) $\gamma_1/\gamma_0=0.9$, (c) $\gamma_1/\gamma_0=1$, (d) $\gamma_1/\gamma_0=1.005$, (e) $\gamma_1/\gamma_0=1.5$, (f) $\gamma_1/\gamma_0=3$. (a),(b) For $\gamma_1<\gamma_0$ there exists two Type B Dirac lines (red and blue) extending through the whole Brillouin zone along the $k_z$ direction. The light red and light blue curves show replicas of these two Dirac lines which are obtained by shifting the original Dirac lines with a reciprocal lattice vector.  (c) At $\gamma_1=\gamma_0$ there occurs a topological phase transition where the Type B Dirac lines merge in a pairwise manner and turn into Type A Dirac lines. The merging of the Type B Dirac lines occur at the time-reversal invariant momenta $\mathbf{k}^{1,2,3}_T=:(0,  2 \pi/\sqrt{3},  -2\pi/3)$, $(\pi, \pm \pi/\sqrt{3}, \pm 2\pi/3)$ (intersection points of red and blue lines). (d), (e) For $\gamma_1 > \gamma_0$ the system supports a single Type A Dirac line. Replicas of this Dirac line are shown with light red and light blue curves to illustrate how this Type A Dirac line is created by merging the Type B Dirac lines. By increasing $\gamma_1/\gamma_0$ further, the radius of the Type A Dirac line shrinks. (f) At  $\gamma_1 =3 \gamma_0$ Type A Dirac line  shrinks to a point, and there occurs another topological phase transition where the system becomes fully gapped by increasing $\gamma_1/\gamma_0$ further.
We have assumed $\gamma_3=0$ but this parameter only renormalizes the critical points of the topological phase transitions so that the picture stays qualitatively the same as long as $|\gamma_3|<0.5 \gamma_0$. The other further neighbor hoppings $\gamma_2$ and $\gamma_4$ do not have any effect on the Dirac lines.}
\label{fig:top_phase_transition}       
\end{figure}

Typical realization of stacked honeycomb lattices is one where the layers are loosely coupled to each other (such as graphite). In that case the band structure parameters obey the hierarchy  $\gamma_0 \gg \gamma_1 \gg \gamma_2, \gamma_3, \gamma_4$ and the model for the rhombohedrally stacked honeycomb lattices supports a pair of Dirac lines in the momentum space extending through the whole Brillouin zone in the $k_z$ direction \cite{Heikkila-Volovik11}. These Dirac spirals are centered at the $K$ and $K'$ points in the $(k_x, k_y)$-plane, so that the projection of the spiral into this plane is a circle with radius proportional to $\gamma_1/\gamma_0$ [see Fig.~\ref{fig:top_phase_transition}(a),(b)]. As discussed in the previous section these Type B Dirac lines are protected by SU(2) spin rotation, time-reversal and inversion symmetries, and they can be gapped only by merging them in a pairwise manner. We first concentrate on the evolution of these Dirac lines with increasing $\gamma_1/\gamma_0$ for $\gamma_2=\gamma_3=\gamma_4=0$, and discuss the effects of the further-neighbor hoppings $\gamma_2, \gamma_3$ and $\gamma_4$ afterwards.

The evolution of the Dirac lines with increasing $\gamma_1/\gamma_0$ is shown in Fig.~\ref{fig:top_phase_transition}. For $\gamma_1<\gamma_0$ there exists two Type B Dirac lines (red and blue) extending through the whole Brillouin zone along the $k_z$ direction [Fig.~\ref{fig:top_phase_transition}(a),(b)]. At $\gamma_1=\gamma_0$ there occurs a topological phase transition where the Type B Dirac lines merge in a pairwise manner and turn into Type A Dirac lines [Fig.~\ref{fig:top_phase_transition}(c)]. The merging of the Type B Dirac lines occur at the time-reversal invariant momenta $\mathbf{k}^{1,2,3}_T=:(0,  2 \pi/\sqrt{3},  -2\pi/3)$, $(\pi, \pm \pi/\sqrt{3}, \pm 2\pi/3)$. In the vicinity of these merging points at the topological phase transition ($\gamma_1 = \gamma_0$) the low-energy theory can be written as 
\begin{equation}
H(\mathbf{k}^1_T+\mathbf{q})= \gamma_0 \bigg(q_z-\frac{2}{\sqrt{3}}q_y\bigg) \sigma_y+\gamma_0 \bigg(\frac{-3 q_x^2+q_y^2+6 q_z^2}{12} \bigg) \sigma_x \label{BA-trans}
\end{equation}
This means that in the ($k_x, k_y$)-plane the Dirac dispersions turns into a semi-Dirac dispersion at the topological phase transition. Somewhat similar topological phase transitions where the merging of two Dirac points leads to an appearance of semi-Dirac fermions have been studied previously in various two-dimensional systems \cite{Horsdal17, Hasegawa06, Katayama06,Montambauz09, Tarruell12, Bellec13, Rechtsman13, Duca15, Kim-merging-15} and topological phase transitions associated with merging of Weyl points have also been studied in three-dimensional systems \cite{Volovi07, Murakami07, Murakami08}. Generically, at the merging transitions the dispersion turns from linear to parabolic in the direction where the merging occurs. In this case the dispersions around the Dirac lines are linear in $q_x$ before the merging ($\gamma_1<\gamma_0$) but they turn parabolic in $q_x$ at the transition ($\gamma_1 = \gamma_0$).

For $\gamma_0 <\gamma_1 < 3 \gamma_0$ the system supports a single Type A Dirac line [Fig.~\ref{fig:top_phase_transition}(d),(e)].  By increasing $\gamma_1/\gamma_0$, the radius of the Type A Dirac line shrinks. At  $\gamma_1 =3 \gamma_0$ the Type A Dirac line shrinks to a point at the time-reversal invariant momentum $\mathbf{k}^{4}_T=(0,0,\pi)$, and there occurs another topological phase transition where the system becomes fully gapped by increasing $\gamma_1/\gamma_0$ further. In the vicinity of the transition the low-energy theory can be written as 
\begin{equation}
H(\mathbf{k}^4_T+\mathbf{q})= \gamma_1 q_z \sigma_y+ \bigg(3 \gamma_0-\gamma_1+\frac{-\gamma_0 q_x^2-\gamma_0 q_y^2+2 \gamma_1 q_z^2}{4} \bigg) \sigma_x. \label{Agapped-trans}
\end{equation}
This low-energy Hamiltonian describes the typical phase transition  where as a function of a parameter $\gamma_1$ the radius of Type A Dirac loop (proportional to $\sqrt{3 -\gamma_1/\gamma_0}$ for $\gamma_1<3\gamma_0$) shrinks to a point at the transition ($\gamma_1=3 \gamma_0$) and then the system becomes fully gapped by further increasing the parameter ($\gamma_1 >3 \gamma_0$). Similar transition is discussed for example in Ref.~\cite{Fu15} for Dirac loops which do not support a monopole charge so that they can be gapped after they have been shrunk to a point. 

We now discuss the effects of the further neighbor hoppings on the transitions discussed above. The terms $\gamma_2$ and $\gamma_4$ do not have any effect on the Dirac lines or transition points except that they shift the band crossings to a finite energy. The terms proportional to $\gamma_2$ and $\gamma_4$ give rise to extra terms in the low-energy Hamiltonians  (\ref{BA-trans}) and  (\ref{Agapped-trans}) 
\begin{equation}
\delta H(\mathbf{q})= (C+D_x q_x^2+D_y q_y^2+D_z q_z^2+E q_y q_z) \sigma_0,
\end{equation}
where $C$, $D_x$, $D_y$, $D_z$ and $E$ are coefficients which depend on $\gamma_2$ and $\gamma_4$. These terms only lead to small quantitative changes in the dispersions around the band crossing points but do not affect the qualitative nature of the transitions. 

The parameter $\gamma_3$ only renormalizes the critical points of the topological phase transitions, but the picture stays qualitatively the same and the transitions even occur at the same time-reversal invariant momenta $k_T^{1,2,3,4}$ as long as $|\gamma_3|<0.5 \gamma_0$. In the presence of $\gamma_3$ the topological phase transition from Type B Dirac lines to Type A Dirac lines occurs at $\gamma_1=\gamma_0+\gamma_3$ and the transition from Type A Dirac line to a gapped system occurs at $\gamma_1=3\gamma_0-3\gamma_3$. Additionally, $\gamma_3$ just renormalizes the numerical coefficients in the low-energy Hamiltonians and gives rise to some unimportant cross terms proportional to $q_y q_z$ multiplying $\sigma_x$ in the low-energy Hamiltonian (\ref{BA-trans}). It does not lead to terms proportional to $q_x q_z$ and $q_x q_y$ because of the mirror symmetry of the structure with respect to the $(y,z)$-plane. In the Hamiltonian (\ref{Agapped-trans}) these cross terms cannot arise at all because of the mirror symmetry with respect to the $(y,z)$-plane and rotation symmetries by $\pm 2\pi/3$ around the $z$-axis.

\section{Surface state spectrum}
\label{sec:surface_states}

\begin{figure}
\begin{center}
  \includegraphics[width=0.99\textwidth]{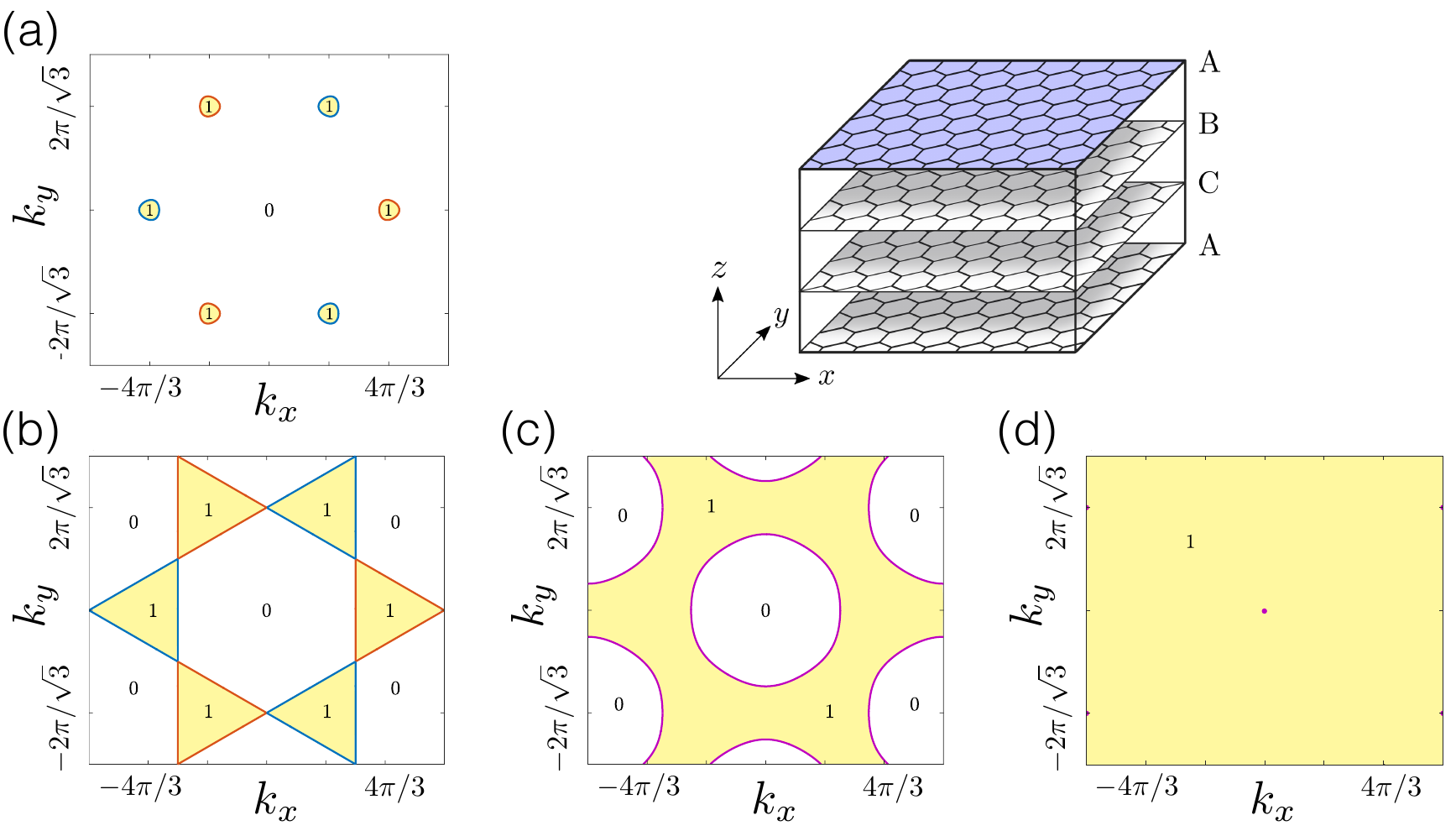}
  \end{center}
\caption{$W(k_x,k_y)$ for $\gamma_2=\gamma_3=\gamma_4=0$ and various values of $\gamma_1/\gamma_0$: (a) $\gamma_1/\gamma_0=0.3$, (b) $\gamma_1/\gamma_0=1$, (c) $\gamma_1/\gamma_0=1.5$, (d) $\gamma_1/\gamma_0=3$. $|W(k_x,k_y)|$ determines the number of zero-energy surface states at the top and bottom surfaces. The transitions between different values of $W$ occur at  the momenta of the projected Dirac lines so that the surface states form flat bands in regions of $(k_x, k_y)$ bounded by the projected Dirac lines. (a) For $\gamma_1<\gamma_0$ there exists two Type B Dirac lines and the flat bands are formed inside the regions bounded by each of them separately (regions bounded by red and blue lines).  (b) At $\gamma_1=\gamma_0$ there occurs a topological phase transition where Type B Dirac lines merge in a pairwise manner and turn into Type A Dirac lines. This merging transition shows up as a unification of the regions of the flat bands. (c) For $\gamma_1 > \gamma_0$ the system supports a single Type A Dirac line. The flat band is now formed everywhere outside the region bounded by its projection. (d) At  $\gamma_1 =3 \gamma_0$ the Type A Dirac line shrinks to a point and surface flat bands now appear at all momenta $(k_x, k_y)$ except this point $(k_x, k_y)=\vec{0}$.}
\label{fig:surfacestates_top}       
\end{figure}

In order to compute the momentum-space structure of the surface states we follow a similar approach as in Ref.~\cite{nexus_Hyart}. Namely, we first assume that the further neighbor hoppings can be neglected $\gamma_2=\gamma_3=\gamma_4=0$, and discuss their effects afterwards. The Hamiltonian then satisfies a chiral symmetry [Eq.~(\ref{chiral-symmetry})], which simplifies the calculation of the surface state spectrum. We start by considering the surface states at the top and bottom surfaces (Fig.~\ref{fig:surfacestates_top}). Because of translational invariance in the $x$- and $y$-directions $k_x$ and $k_y$ are good quantum numbers and by fixing them we get a 1D Hamiltonian $H_{k_x, k_y}(k_z)$, which depends only on $k_z$. Because of the chiral symmetry the 1D Hamiltonians $H_{k_x, k_y}(k_z)$ have well-defined topological invariants. To calculate this topological invariant, we first notice that the Hamiltonian can be written in a block-off-diagonal form
\begin{equation}
H(\mathbf{k})=\begin{pmatrix}
0 & \Phi(\mathbf{k}) \\
\Phi^*(\mathbf{k}) & 0
\end{pmatrix}. \label{Block-off}
\end{equation}
The topological invariant can then be defined as a winding number
\begin{eqnarray}
W(k_x, k_y)=-\frac{i}{2 \pi} \int \frac{dz(k_z)}{z}, \ z=\frac{\Phi(\mathbf{k})}{|\Phi(\mathbf{k})|},
 \label{W-supp}
\end{eqnarray} 
where the integration is over the 1D Brillouin zone in $k_z$ direction. $W(k_x, k_y)$ is well-defined whenever there are no gap closings as a function of $k_z$, and in these cases $|W(k_x, k_y)|$ determines the number of zero-energy surface states for each $k_x$ and $k_y$. Therefore, the winding number and the number of topologically protected zero-energy states can only change at the momenta of the projected Dirac lines. By computing $W(k_x, k_y)$ for various values of $\gamma_1/\gamma_0$ we arrive at a flat band (zero energy) spectrum in the regions of the $(k_x, k_y)$ with $W\ne 0$ in Fig.~\ref{fig:surfacestates_top}.  For $\gamma_1<\gamma_0$ the flat bands are formed inside the regions bounded by each of the Type B Dirac lines separately [Fig.~\ref{fig:surfacestates_top}(a)] as discussed in Refs.~\cite{Heikkila-Volovik11, Heikkila-flat-bands}. 
 At $\gamma_1=\gamma_0$ the Type B Dirac lines merge in a pairwise manner and turn into Type A Dirac lines, and this shows up as a unification of the regions of the flat bands [Fig.~\ref{fig:surfacestates_top}(b)]. Similar merging of flat band surface states in a rhombohedral multilayer graphene-like system has been considered earlier as a function of an increasing anisotropy in the intra-layer hopping parameters \cite{Zyuzin15}. Interestingly, for $\gamma_1 > \gamma_0$ the flat band is formed everywhere outside the region bounded by the projected Type A Dirac line [Fig.~\ref{fig:surfacestates_top}(c)], and therefore when the Type A Dirac line has shrunk to a point at $\gamma_1 =3 \gamma_0$ the surface flat bands appear at all momenta $(k_x, k_y)$ except this point $(k_x, k_y)=\vec{0}$ [Fig.~\ref{fig:surfacestates_top}(d)]. For  $\gamma_1 > 3 \gamma_0$  the system is fully gapped, but the zero-energy surface flat bands still appear at all momentum and they are completely detached from the bulk bands. This type of surface flat bands has been found previously in two-dimensional systems \cite{Ezawasurfaceflatbands}.  

\begin{figure}
\begin{center}
  \includegraphics[width=0.99\textwidth]{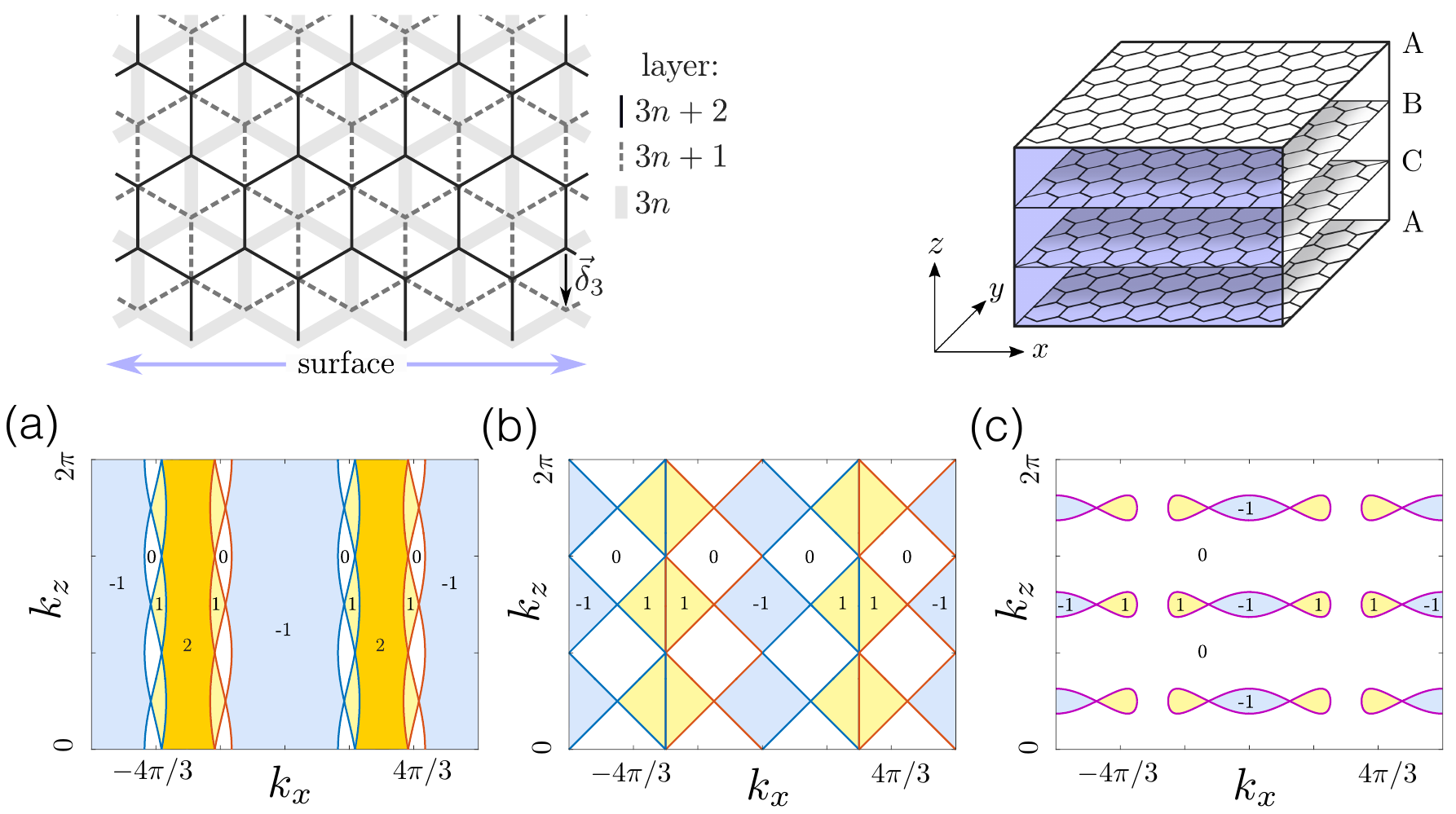}
  \end{center}
\caption{$W(k_x,k_z)$ for $\gamma_2=\gamma_3=\gamma_4=0$ and various values of $\gamma_1/\gamma_0$: (a) $\gamma_1/\gamma_0=0.3$, (b) $\gamma_1/\gamma_0=1$, (c) $\gamma_1/\gamma_0=1.5$. $|W(k_x,k_z)|$ determines the number of zero-energy surface states at the side surface with the specific surface termination shown in the figure both from above and from a three-dimensional perspective. For $\gamma_1<\gamma_0$ there are regions in the momentum space with more than one surface flat band $|W(k_x, k_z)|>1$. Such regions are possible because the period of the surface in the $z$-direction is three times the distance between the layers, so that there exists three different projections of each of the bulk Dirac lines in the $(k_x, k_z)$-plane (red and blue lines). After the merging transition ($\gamma_1 > \gamma_0$) the Type A Dirac line gives rise to regions with $W(k_x, k_z)=1$ and $W(k_x, k_z)=-1$. The different signs of $W(k_x, k_z)$ indicate that the surface states are localized in the different sublattices. By increasing $\gamma_1/\gamma_0$ the regions of the flat bands shrink and they disappear at $\gamma_1 = 3 \gamma_0$.}
\label{fig:surfacestates_side}       
\end{figure}

We can now use a similar procedure to compute the momentum-space structure of the surface states at the side surfaces. This way we obtain that $W(k_y, k_z)=0$ for all values of $k_y$ and $k_z$, which means that there are no surface flat bands at the surface perpendicular to the $x$-direction. This result is analogous to the earlier finding that there are no flat bands at the armchair edge in graphene \cite{Ryu02}.  (The surface perpendicular to the $x$-direction consists of a stack of armchair edges.) On the other hand, for the surface perpendicular to the $y$-direction we find regions with $W(k_x, k_z)\ne 0$ indicating the existence of flat bands. The momentum-space structure flat bands for various values of $\gamma_1/\gamma_0$ are shown in Fig.~\ref{fig:surfacestates_side}. Importantly, for $\gamma_1<\gamma_0$ there exist also regions in the momentum space with more than one surface flat band $|W(k_x, k_z)|>1$. Such regions are possible because the period at the surface in the $z$-direction is three times the distance between the layers, and therefore there are three different projections of each of the bulk Dirac lines in the $(k_x, k_z)$-plane (see Fig.~\ref{fig:surfacestates_side}).  The detailed momentum-space structure of the flat bands for $\gamma_1<\gamma_0$ can be understood by noticing that the surface perpendicular to the $y$-direction contains a periodic sequence of stacks with two zigzag edges and one bearded edge (see Fig.~\ref{fig:surfacestates_side}). Therefore, by utilizing the results found in Ref.~\cite{Ryu02}, we can interpret our numerical results in such a way that the two zigzag edges essentially give rise to the two flat bands in the region with $W(k_x, k_z)=2$ and the bearded edge is responsible for the region with $W(k_x, k_z)=-1$  in Fig.~\ref{fig:surfacestates_side}(a). The different signs of $W(k_x, k_z)$ indicate that the surface states are localized in the different sublattices. After the merging transition ($\gamma_1 > \gamma_0$) the Type A Dirac line gives rise to regions with $W(k_x, k_z)=1$ and $W(k_x, k_z)=-1$ indicating that there still exists surface states localized in the different sublattices. By increasing $\gamma_1/\gamma_0$ the regions of the flat bands shrink and they disappear at the topological phase transition where the system becomes gapped ($\gamma_1 = 3 \gamma_0$). The detailed momentum-space structure of the flat bands depends on exact surface termination similarly as in the case of graphene \cite{Ryu02}. The only generic property of the surface states, which is independent of the surface termination, is that the surface flat bands are always bounded in the momentum space by the projected Dirac lines. 

The effects of the further neighbor hoppings on the surface states can be computed quantitatively similarly as in Ref.~\cite{nexus_Hyart} for the case of Bernal graphite. Here, we only discuss these effects qualitatively. The term proportional to $\gamma_3$ obeys chiral symmetry. Therefore, the same procedure of calculation of $W(k_x, k_y)$ and $W(k_x, k_z)$ can be repeated also for $\gamma_3 \ne 0$. Because for $|\gamma_3| < 0.5 \gamma_0$ all the transitions stay qualitatively similar, the only effect of $\gamma_3$ in this regime is a small modification of the shape of the projected Dirac lines, so that the boundaries of the regions where the flat bands appear are slightly modified. On the other hand, the parameters $\gamma_2$ and  $\gamma_4$ break the chiral symmetry and cause the Dirac lines to appear at finite energy. Similarly as found in Ref.~\cite{nexus_Hyart} for the case of Bernal graphite, the terms breaking the chiral symmetry also modify the dispersions of the surface states, so that they turn the flat bands into a drumhead surface which are bounded by the projected Dirac lines both in energy and momentum. In the region in between the Dirac lines the surface states are no longer flat and their dispersions are determined by the terms proportional to $\gamma_2$ and $\gamma_4$ \cite{KopninPRB}. If these parameters are small the surface bands are still approximately flat.

\section{Summary and discussion}

We have shown that in the presence of time-reversal, inversion and spin rotation symmetries there can exist two different types of Dirac lines depending on whether the Dirac lines form closed loops fully inside the Brillouin zone (Type A Dirac lines) or they extend through the whole Brillouin zone in one of the directions (Type B Dirac lines). In the case of Type B Dirac lines, an energy gap can be opened only by first merging the Type B Dirac lines in a pairwise manner so that they turn into Type A Dirac lines, and then by shrinking these loops into points. We show that this kind of topological phase transition can occur in rhombohedrally stacked honeycomb lattices by tuning the ratio of the tunneling amplitudes.  We have also discussed the properties of the surface states in the different phases of the model.

The Dirac line semimetals considered in this paper are particularly interesting because of the possible symmetry-broken states at the surface  triggered by the large density of states caused by the topologically protected approximately flat bands. These symmetry-broken states may for example lead to realization of high-temperature superconductors or interesting magnetic orders \cite{Heikkila11b, Mauri17, Black-Schaffer17}. Moreover, they are expected to be exotic states of matter since they cannot be described with a mean field theory \cite{Kauppila}.

Rhombohedral graphite has been studied also experimentally, and evidence of the surface flat bands has been reported \cite{Pierucci15, Henni16}.
Moreover, the accumulated experimental evidence (e.g. sharp drop of resistance as a function of temperature and Josephson-like I-V characteristic) for graphite samples indicates the existence of high-temperature granular superconductivity which is localized at internal interfaces \cite{Esquinazi08, Scheike12, Ballestar13, Precker16, Stiller17}. Nevertheless, a development of a consistent and comprehensive theory for all graphite experiments is a difficult and unsolved problem. In particular, the recent experiments for rhombohedral graphite were interpreted as evidence of a bulk energy gap $E_{\rm gap} \sim 100$ meV for rhombohedral graphite \cite{gapingraphite}.  In the light of the theory developed in this paper this observation is mysterious. Namely, according to the theory the rhombohedral graphite should be a Type B Dirac-line semimetal protected by the lattice translation, time-reversal, inversion and spin rotation symmetries. Moreover, the only way to open an energy gap in the presence of these symmetries (in a mean field theory description) is to merge the two Type B Dirac lines with each other, which requires a huge perturbation on the order of $\gamma_0$. Since this is not feasible in rhombohedral graphite the only possible explanation (assuming high-quality crystal structure so that the symmetries are not explicitly broken) would be that this bulk energy gap is due to interaction effects. Strong interactions could in principle either lead to a spontaneous symmetry-breaking in the bulk destroying the protection of the Dirac lines or to a strongly correlated state which cannot be described with a mean field theory. In both cases the opening of an energy gap is in principle possible. Moreover, such kind of strong interaction effects are in principle possible. According to a simple theoretical estimate the Hubbard $U$ parameter in graphite can be $U \sim 6$ eV \cite{Hubbard-U-graphite}, whereas $\gamma_0$ is typically assumed to be $\gamma_0 \sim 2.8$ eV \cite{graphene-review}. 

In addition to the relevance of our theory for the rhombohedral graphite, we have made explicit predictions concerning the topological phase transition between Type B and Type A Dirac-line semimetals. These transitions could be realized experimentally in systems where the lattice potential can be controlled, so one essentially needs a three-dimensional generalizations of the type of two-dimensional systems where the merging transition of Dirac points have already been observed \cite{Tarruell12, Bellec13, Rechtsman13,Duca15}. 

\begin{acknowledgements}
We thank G.~E.~Volovik, T.~Bzdusek and A. Bouhon for fruitful discussions and comments. This work was supported by the Academy of Finland Centre of Excellence and Key Funding programs (projects No. 284594 and 305256).

\end{acknowledgements}


\begin{thebibliography}{99}
\bibitem{comment}
Berry phase $\phi$ depends on the convention used for the overall phase of the wavefunctions. Therefore, it is uniquely defined only up to $n 2 \pi$ ($n \in \mathbb{Z}$). Here, for simplicity we fix the convention for the overall phase in such a way that $0 \leq \phi < 2 \pi$. This automatically fixes also a specific convention for the Berry connection $\vec{\cal{A}}(\mathbf{k})$.
\bibitem{comment2}
We can generalize the argument also to the case where the system has a chiral symmetry. In this case the Hamiltonian can be always block off-diagonalized and the Berry phase $\phi/\pi$ in Eq.~(\ref{monopole-charge}) can be replaced with the winding number of the determinant of the off-diagonal block of the Hamiltonian. The difference is that this new winding number does not have the same ambiguity as the Berry phase related to the shifts of $n 2 \pi$ ($n \in \mathbb{Z}$) and therefore $Q_M$ becomes a $\mathbb{Z}$ topological invariant. Because in these symmetry classes $|Q_M|=\nu$ only the Dirac lines with $\nu=0$ can be created and annihiliated individually. In the special case of 2$\times2$ Hamiltonian with time-reversal and inversion symmetries, the Hamiltonian automatically has a chiral symmetry up to terms proportional to $\sigma_0$. Because the terms proportional to $\sigma_0$ do not influence the existence of the Dirac lines the $\mathbb{Z}$ topological invariant can be defined also in this case.
\end{thebibliography}
\end{document}